\documentclass[sigconf, screen]{acmart}
\AtBeginDocument{%
  }

\usepackage{multirow}
\usepackage{graphicx}
\usepackage{tabularx, booktabs, array, rotating}
\usepackage{microtype}

\setcopyright{acmlicensed}
\copyrightyear{2026}
\acmYear{2026}
\acmDOI{XXXXXXX.XXXXXXX}
\acmISBN{978-1-4503-XXXX-X/2018/06}

\begin{document}

\title[Exploring and Analyzing the Effect of Avatar's Visual Style ...]{Exploring and Analyzing the Effect of Avatar's Visual Style on Anxiety of English as Second Language (ESL) Speakers}


\author{Tian-Qi Liu}
\affiliation{%
  \institution{Department of Information Science, \\Cornell University}
  \city{Ithaca}
  \country{USA}
}
\email{tl889@cornell.edu}

\author{Xin Yi}
\affiliation{%
  \institution{Institute for Network Sciences and Cyberspace, \\Tsinghua University}
  \city{Beijing}
  \country{China}}
\email{yixin@tsinghua.edu.cn}

\author{Yuan-Chun Shi}
\authornote{Corresponding author}
\affiliation{%
  \institution{Department of Computer science and Technology, \\Tsinghua University}
  \city{Beijing}
  \country{China}
}
\email{shiyc@tsinghua.edu.cn}

\author{Yun-Tao Wang}
\authornote{Corresponding author}
\affiliation{%
  \institution{Department of Computer Science and Technology, \\Tsinghua University}
  \city{Beijing}
  \country{China}
}
\email{yuntaowang@tsinghua.edu.cn}

\renewcommand{\shortauthors}{Liu et al.}

\begin{abstract}
Virtual avatars offer new opportunities to reshape communication experiences beyond traditional live video. However, it remains unclear how avatar representations influence communication anxiety for English as a Second Language (ESL) speakers, and why such effects emerge.

To take a first step to address this, we conducted a controlled laboratory study in which Mandarin-speaking ESL participants engaged in one-on-one conversations under three representation conditions: live video, stylized avatars, and realistic avatars. We assessed anxiety using both self-reported measures and physiological signals (EDA, ECG, PPG).

Our results show that avatar style plays a critical role in shaping communication anxiety. While live video remained a strong baseline with low subjective anxiety, stylized avatars achieved comparable — and in some cases lower — physiological anxiety levels, whereas realistic avatars elicited higher anxiety. Beyond these effects, our findings reveal three underlying mechanisms that explain how avatar representations shape ESL communication anxiety: (1) facial expressiveness; (2) perceived feedback and fear of negative evaluation; and (3) contextual appropriateness.

This work provides actionable design implications for developing avatar-mediated communication systems that support emotionally sustainable cross-linguistic interaction.
\end{abstract}




\maketitle
\section{Introduction}

Screen-based communication has become an increasingly important medium for cross-cultural interaction~\cite{koyano2022development, avatarEducationPandemic}. 
Live video is commonly regarded an effective approach to support remote communication because it provides rich visual cues and supports social presence between interlocutors~\cite{pavlov2021beyond, powers2013conversations}. 
However, concerns about privacy of using live video and the opportunity of reshaping experience have motivated the exploration of alternatives such as virtual avatars to camera-based interaction.

Avatars can substantially reduce privacy concerns while also reshaping how communication partners are visually represented and perceived~\cite{nowak2018avatars, pan2015virtual, gu2025beyond}, by replacing real camera feeds. Although some of the previous work argue that real people lead to a lower fear of communication compared to the virtual avatar~\cite{powers2013conversations}, certain groups—including individuals with anxiety disorders and autistic adults—may prefer avatar-mediated communication over live video. This preference has been attributed to benefits such as masking and reduced social exhaustion~\cite{yoneyama2023augmented, do2025exploring}.
However, they compared a single avatar representation with live video, leaving it unclear how different avatar visual styles influence communication anxiety. Overall, these findings suggest that avatar-mediated communication can provide valuable opportunities in contexts where psychological factors play an important role. 

An important such context is communication for English as a Second Language (ESL) speakers. 
Communication anxiety is widely recognized as a major barrier affecting ESL speakers' confidence, engagement, and willingness to participate in English conversations~\cite{song2024foreign}. 
This form of anxiety is not driven solely by linguistic proficiency; it is also strongly influenced by social cues and psychological factors such as perceived evaluation, fear of negative evaluation, and uncertainty about making mistakes during interaction~\cite{tanveer2007investigation, horwitz1986foreign}. 
Consequently, the visual representation of interlocutors may play a meaningful role in shaping ESL speakers' communication experiences.

\textit{Visual style} is one of the most fundamental design dimension of avatars, including stylized representations with exaggerated or cartoon-like features and realistic representations with human-like appearance~\cite{canales2024impact}.
Previous work has shown that avatar visual style can influence users' experience from different psychological aspects. 
For example, stylized avatars may evoke affinity in certain contexts~\cite{inkpen2011me}, whereas realistic virtual characters may increase perceived enjoyment, but can also evoke discomfort associated with the uncanny valley phenomenon~\cite{rogers2022realistic, geller2008overcoming}. 


Prior work has primarily focused on user experience outcomes such as affinity or eeriness, offering limited insight into how these perceptual differences translate into communication anxiety. This gap is particularly important in ESL contexts, where anxiety is a dominant factor influencing communication engagement.

In addition, existing research lacks a mechanistic understanding of how and why different avatar styles shape communication anxiety. Furthermore, prior studies have predominantly relied on self-reported measures, leaving it unclear whether these effects are also reflected in users’ physiological responses. Such measures can provide complementary evidence of underlying affective processes, enabling a more comprehensive understanding of communication anxiety.

To address these gaps, we investigate how stylized and realistic avatars influence ESL speakers’ communication anxiety, in comparison to live video interaction. We assess anxiety using both subjective self-reports and physiological signals, enabling a more comprehensive understanding of users’ responses.

Our results show that stylized avatars reduce communication anxiety relative to realistic avatars and, from a physiological perspective, can match and slightly outperform live video. We further identify three key mechanisms—facial expressiveness, perceived feedback and fear of negative evaluation, and contextual appropriateness—that help explain how avatar representations shape communication anxiety. These findings suggest that stylized avatars may provide a promising design direction for avatar-mediated communication systems aimed at supporting ESL speakers.

\section{Related Work}
Previous work related to this study spans three main research threads: the effects of avatar style on user experience, anxiety in ESL communication, signal processing methods to detect anxiety.

\subsection{Avatar Visual Style and It's Effect on Social Experience}
\subsubsection{Stylized Avatar and Realistic Avatar}
\label{defination of stylized avatar}

\begin{figure}[ht]
\centering
\includegraphics[width=\columnwidth]{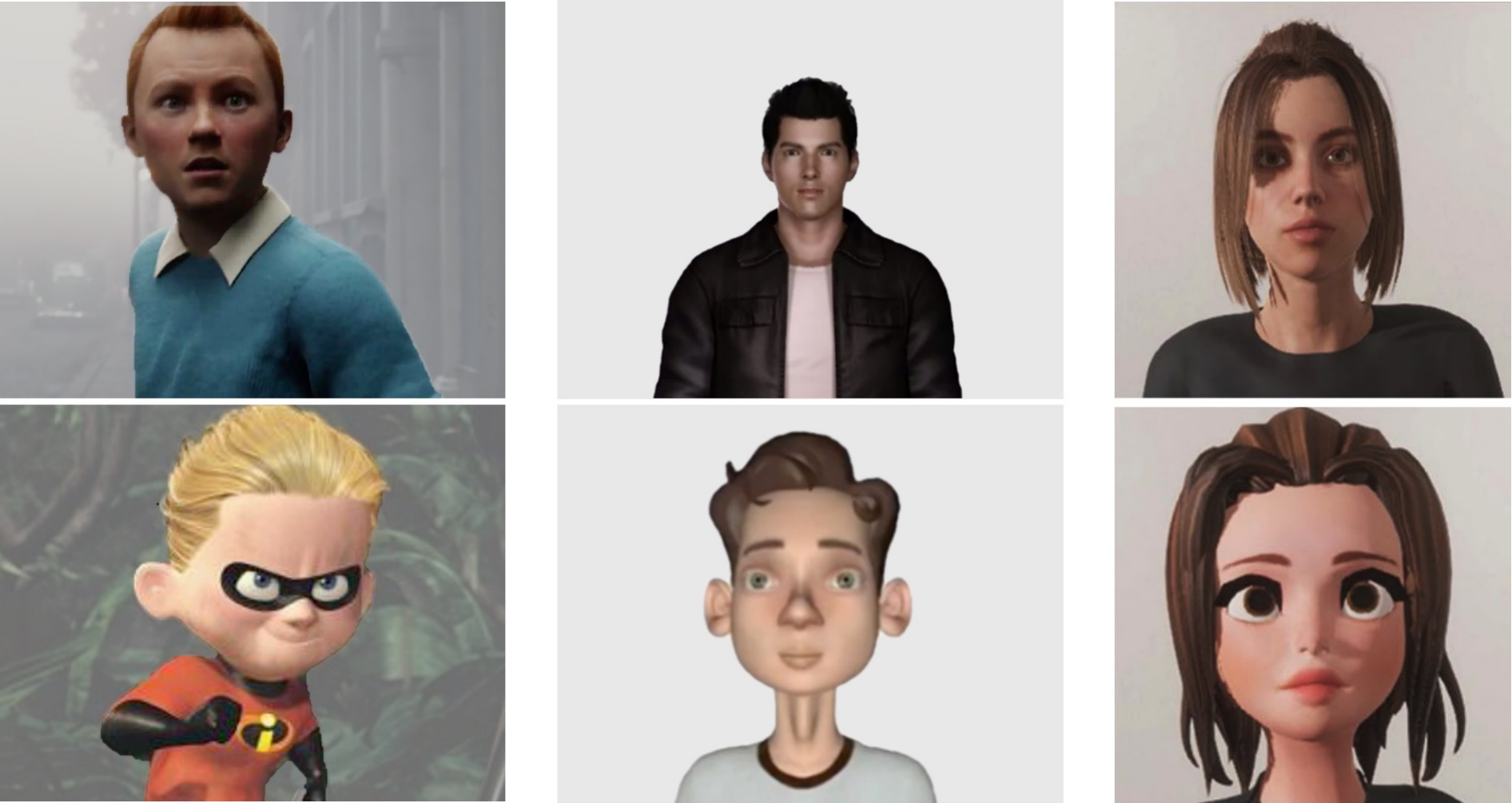}
\caption{Examples of realistic and stylized avatars used in prior work. 
Left: reference images illustrating realistic (top) and stylized (bottom) avatars in~\cite{adamo2016toward}. 
Middle: realistic (top) and stylized (bottom) avatars designed in~\cite{canales2024impact}.
Right: realistic (top) and stylized (bottom) avatars designed in~\cite{dubosc2025effect}.
}
\label{fig_avatars_related_works}
\end{figure}

Stylization refers to variations in the visual appearance of avatars across different design styles. 
Prior work commonly distinguishes between \textit{stylized avatars}, which adopt cartoon-like visual features, and \textit{realistic avatars}, which aim to approximate human appearance~\cite{canales2024impact, dubosc2025effect}. 

Figure~\ref{fig_avatars_related_works} illustrates representative examples of stylized and realistic avatars used in prior research. Stylized avatars often employ simplified geometry and exaggerated facial features, including larger head-to-body ratio, rounded facial shapes, and enlarged eyes. Their design is commonly inspired by animation styles such as those used in Disney or Pixar characters~\cite{dubosc2025effect, dubosc2025effect}. 
Realistic avatars attempt to resemble real humans as closely as possible, often preserving facial proportions and visual details derived from 2D photographs~\cite{makarainen2014exaggerating} or 3D scans~\cite{canales2024impact}.


\subsubsection{Avatar Visual Style Effect on Social Experience}
Previous research shows that the visual style of virtual avatars can shape user interaction experiences through multiple psychological mechanisms. For example, stylized avatars with exaggerated features (e.g., larger eyes) may evoke a higher affinity due to long-term exposure and adaptation of users to cartoon-like faces in media and entertainment~\cite{chen2010crossing}. Familiarity with cartoon representations may further increase users' empathy when interacting with similar avatars~\cite{dill2012evaluation}. The sense of social co-presence was also reported to be higher when interacting with stylized avatars compared to realistic ones~\cite{dubosc2025effect}. Another study reported that trust was not affected by avatar style~\cite{canales2024impact}. 

Findings regarding more human-like avatar styles are mixed. On one hand, avatars with more human-like appearances can enhance aspects of user experience such as perceived enjoyment during interaction~\cite{rogers2022realistic}. On the other hand, when avatars approach human realism but remain imperfect, they may evoke eeriness—commonly described as the uncanny valley phenomenon~\cite{mcdonnell2012render, social2017Latoschik}. Together, these studies suggest that avatar style can influence user experience in complex ways through different psychological factors such as affinity, presence, enjoyment, and eerieness.

Prior work suggests that human experience in avatar-mediated interaction is shaped by multiple psychological factors. Manipulating avatar visual style (e.g., stylized vs.\ realistic) may simultaneously influence users' perceptions of affinity, eerieness, among other factors. As a result, avatar style can be understood as a bundled perceptual manipulation that affects social experience through co-varying mechanisms. In this work, we examine how such differences in avatar style are reflected in \textit{communication anxiety}, measured through both physiological signals and subjective self-reports.

In this context, it is unknown how avatar style influences communication anxiety, which is a particularly important psychological factor for ESL speakers. Thus, exploring how avatar style affects anxiety in ESL contexts is important. Besides, while avatars are often positioned as an alternative to live video in remote communication, it remains unclear how different avatar styles compare with traditional camera-based communication in shaping users' anxiety. Addressing these gaps can help understand whether avatar-based communication can be used meaningfully as an alternative to video for ESL speakers.

\subsection{Computer-Mediated Anxiety Treatment for ESL Speakers}
Communication anxiety is common in cross-cultural interactions and can negatively affect user' participation, confidence, and overall interaction experience~\cite{mascia2020emotional, steinkopf2021prerequisites}.
Research suggests that ESL speakers' communicative experiences are shaped not only by language proficiency or individual motivation, but also by broader social factors such as social investment, power dynamics, and mutual understanding within interactional contexts~\cite{peirce1995social}.
Previous work suggests that computer-mediated environments can reduce anxiety through low-stakes interaction settings, reducing the immediacy of evaluation, thus encouraging participation~\cite{hajiyeva2024language}. 
Another research showed that structured online collaborative systems that employ role-based interaction and guided participation can help alleviate anxiety by reducing fear of negative evaluation~\cite{chen2018role}.

However, existing approaches for an ESL context primarily emphasize task structure and interaction mechanics, offering limited insight into how virtual representations can help mitigate communication anxiety.

\subsection{Applicable Signal Processing Methods for Anxiety Detection}

Physiological signal processing offers an objective and widely adopted approach to anxiety detection by capturing bodily responses associated with autonomic nervous system activity~\cite{picard2000affective, wagner2005physiological}. 
Among commonly used signals, the electrocardiogram (ECG) has been extensively used to derive characteristics of heart rate and heart rate variability, which have been shown to reliably reflect anxiety and mental stress~\cite{Miranda2014AnxietyDU, Pham2021HeartRV, moshe2021predicting, brosschot2003heart, tivatansakul2015improvement}. 
Electrodermal activity (EDA) is a particularly sensitive indicator of emotional arousal and has been widely used to assess stress and anxiety in both controlled laboratory settings and real-world contexts~\cite{SEQUEIRA200950, liu2018psychological, diagnostics12081794}. 
Photoplethysmography (PPG), due to its non-invasive nature and ease of deployment, has also been commonly adopted as an alternative signal for anxiety and emotion monitoring, with prior work demonstrating its effectiveness in diverse applications~\cite{Jan2019EvaluationOC, perpetuini2021prediction, diagnostics12081794}.

Despite the extensive use of physiological signals for the detection of general anxiety, using such signals to capture communication anxiety during ESL–native speaker interactions remains a relatively unexplored area.

\section{Methods}

\subsection{Participants}

A total of 25 participants completed the study (11 male, 14 female), with ages ranging from 18 to 26. Due to regional constraints, all participants reported Mandarin as their first language, which represents a limitation of the sample.


We used vlidated metrics Common European Reference Framework for Languages (CEFR)~\cite{council2001common} scale to report participants' English proficiency. This scale is commonly used for ESL context~\cite{kassim2023common} and include three dimensions: listening, spoken interaction, and spoken production. Approximately two-thirds of the participants' proficiency is intermediate in all three dimensions, while the remaining one-third was beginners, and few samples was advanced level. 



\subsection{Apparatus and Materials}
\subsubsection{Avatars}

\begin{figure}[ht]
\centering
\includegraphics[width=\columnwidth]{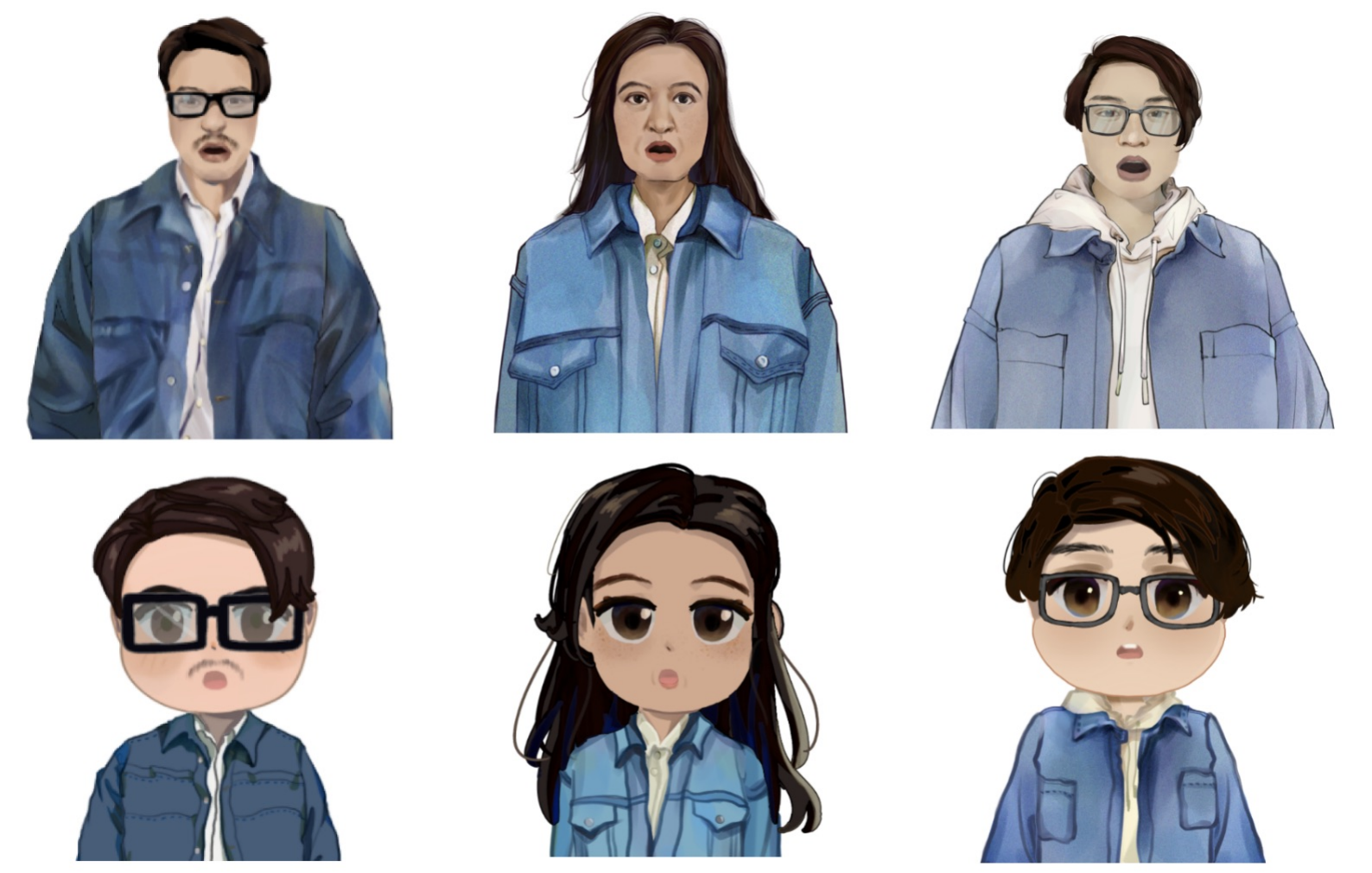}
\caption{Stylized and realistic avatars used in the study. Both avatar styles were custom-designed to reflect the facial features of the corresponding native English speaker.}
\label{fig_avatars}
\end{figure}

Following the definitions of stylized and realistic avatars described in prior work (Section~\ref{defination of stylized avatar}), we designed a total of six avatars based on portrait images of three native English speakers, using Adobe Photoshop (v23.5.5). For each speaker, two avatars were created: one stylized avatar and one realistic avatar. Figure~\ref{fig_avatars} shows the six avatars used in the study.

The stylized avatars were designed with a larger head-to-body ratio, rounded facial shapes, and enlarged eyes~\cite{nam2009understanding, adamo2016toward}. Realistic avatars were designed to approximate the highest level of realism achievable within a 2D representation, following the definition and approach in related work~\cite{makarainen2014exaggerating}, we designed these avatars by tracing facial appearance directly from portrait photographs to preserve realistic facial proportions and visual details.



The avatars were controlled in real time by Adobe Character Animator (v2020), a widely used animation engine for conducting user studies with 2D avatars~\cite{guajardo2024generative, willett2020pose2pose, mellberg20253d}. Facial expressions—including neutral expression, smiling, mouth movements (e.g., opening), blinking, gaze direction, eyebrow movements, and head and upper-body movements—were automatically tracked from the speaker's live video feed and synchronously rendered onto the avatar presented to the participant. 
\subsubsection{Equipment}

Physiological signals, including EDA, ECG, and PPG, were recorded using an MP160 multi-channel physiological recorder. For EDA, two electrodes were placed on the participant's left palm; for PPG, two electrodes were attached to a finger on the left hand; and for ECG, three electrodes were positioned on the participant's chest. The ECG electrodes were attached to the right clavicle and to the ribs on both sides. An Alienware x15 R2 was used for physiological data logging, while a MacBook Pro (2020) served as the primary interface for avatar-mediated conversations.

\subsection{Experimental Design and Procedure}
\begin{figure*}[ht]
\centering
\includegraphics[width=14cm]{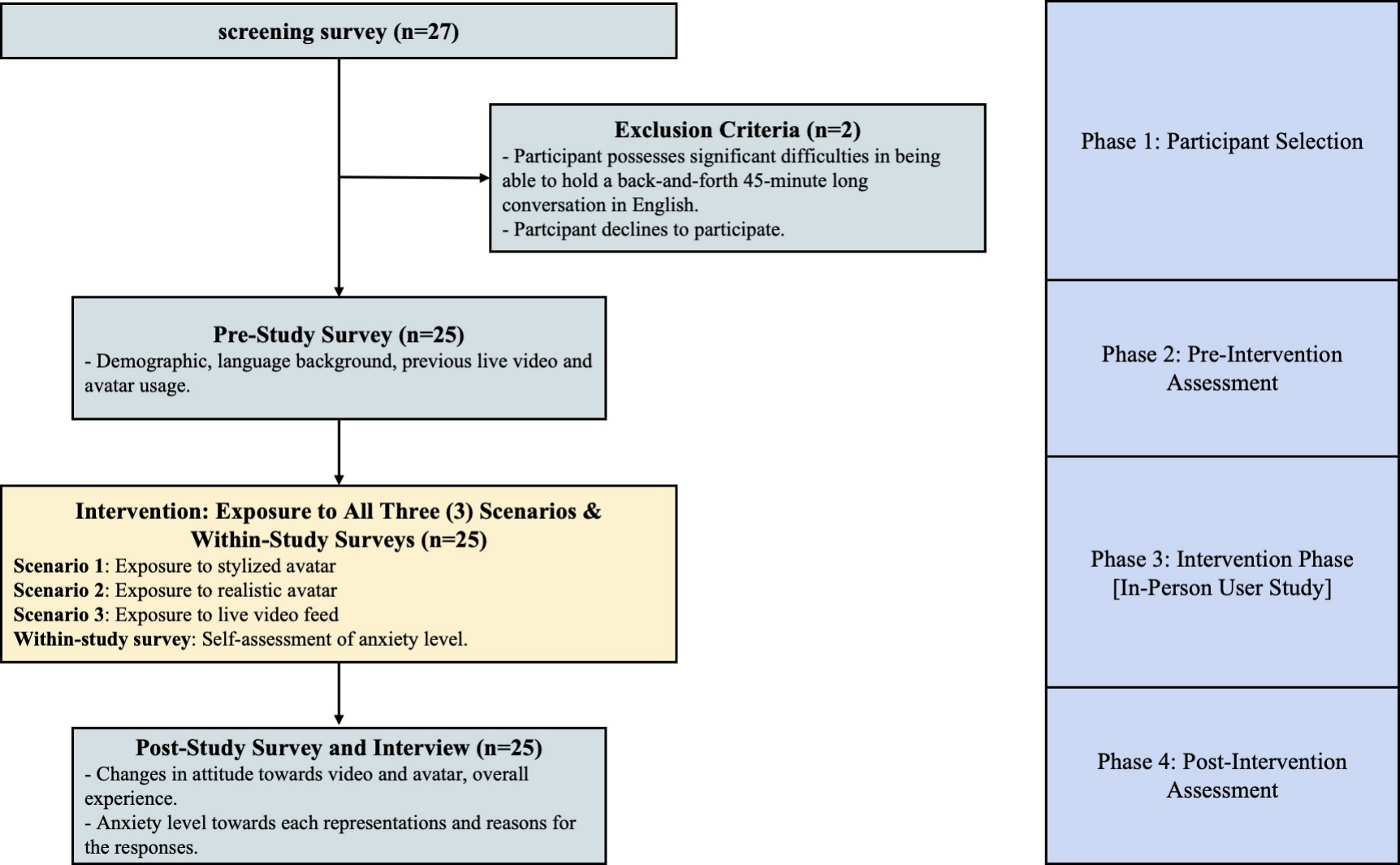}
\caption{Participant Flowchart for the Entire User Study.}
\label{fig_user_study_participant_1}
\end{figure*}

Figure~\ref{fig_user_study_participant_1} illustrates the overall workflow of the user study, including study planning, execution, data collection, and analysis. All study procedures were approved by the institutional review board (IRB). The study consisted of four sequential phases, described in the following.

\textbf{Phase 1: Participant Selection.} Participants were recruited on campus through flyers. Interested individuals completed an online screening survey prior to enrollment. Only English-as-a-Second-Language (ESL) speakers were included. To ensure sufficient communicative ability for the study tasks, individuals who self-reported difficulty maintaining a 45-minute conversation with a native English speaker were excluded.

\textbf{Phase 2: Pre-Intervention Assessment.} 
Before the experimental session, participants completed an online pre-study survey.

\textbf{Phase 3: Intervention Phase.} 
\begin{figure*}[t]
\centering
\includegraphics[width=\linewidth]{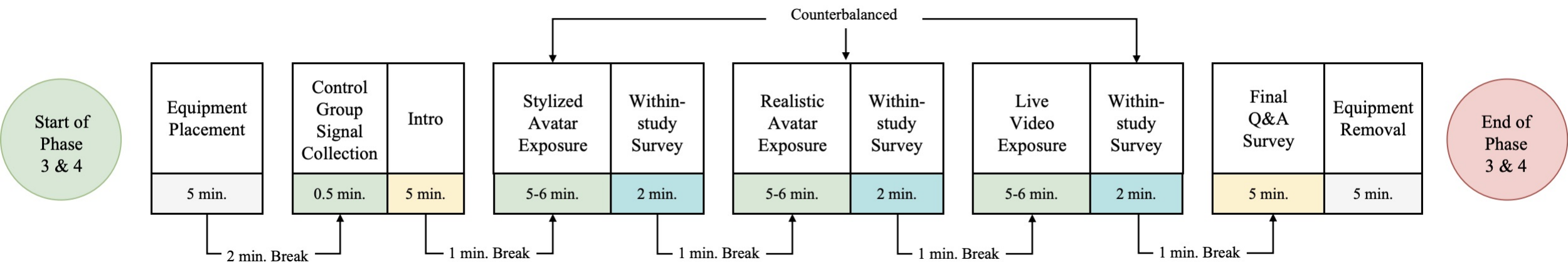}
\caption{Detailed participant flow during the in-person intervention and post-intervention assessment phases.}
\label{fig_user_study_participant_2}
\end{figure*}
The participants visited the laboratory in person. Before the experimental tasks started, the participants were equipped with physiological sensing equipment. After a 2-minute rest period, baseline physiological signals were recorded as the control condition. The participants then received a standardized introduction to the study procedures.

The participants subsequently engaged in three verbal conversations, each with a different native English speaker. Each conversation lasted approximately 5--6 minutes and was conducted under one of three exposure conditions: (1) a stylized avatar, (2) a realistic avatar, or (3) a live video camera stream. Each participant experienced all three conditions exactly once. The order of the conditions was counterbalanced among the participants to control for possible order effects.

To control for confounding variables, we employed an experimental design within-subjects in which the independent variable was the visual representation of the native English speaker (interviewer). All other factors, including the interaction setting, interview duration, and conversation structure, were held constant across conditions. In each exposure, interviewers followed a standardized script consisting of five randomly selected, non-repeating everyday life--related questions designed to prompt natural conversation, following previous works design~\cite{mcdonnell2012render, yoneyama2023augmented}. The Interviewers were instructed to maintain a calm and friendly demeanor throughout all interactions. During the delivery of instructions and logistical explanations unrelated to the experimental exposure, the interviewer's camera was turned off to avoid unintended visual influence.

Immediately after each exposure, the participants completed a brief within-study survey.

\textbf{Phase 4: Post-Intervention Assessment.} After completing all three interaction conditions, participants completed a post-study survey.

Phases 3 and 4 were conducted entirely in the laboratory and together lasted approximately 45 minutes (Figure~\ref{fig_user_study_participant_2}).

\subsection{Data Collection}
\subsubsection{Surveys}

All participants completed four surveys in a fixed sequence: a screening survey, a pre-study survey, a within-study survey administered after each experimental exposure, and a post-study survey. All survey instruments are provided in the Supplementary Materials.

\textbf{Screening Survey.} The screening survey collected participants' contact information and assessed their eligibility for the study, including ESL speaker status and the ability to engage in a 45-minute English-language conversation.

\textbf{Pre-study Survey.} The pre-study survey gathered demographic information, native language background, CEFR English language proficiency.
It also assessed participants' prior experience with using live video and 2D avatars in online communication.

\textbf{Within-study Surveys.} The within-study surveys assessed communication anxiety associated with each visual representation. We employed adapted Likert-scale items designed to capture anxiety in avatar-mediated communication contexts. Because existing validated anxiety scales do not adequately address this interaction setting, we developed study-specific items. The internal consistency of these items was evaluated using Cronbach's alpha, demonstrating good reliability.

\textbf{Post-study Survey and Interview.}
The post-study survey assessed participants' 
changes in their attitudes toward video and avatar use after the study, and collected feedback on their overall study experience.

A follow-up interview further probed participants' perceived communication anxiety across different visual representations, as well as the underlying reasons for their responses.

\subsubsection{Signals}

Three physiological signals, including EDA, ECG, and PPG, were continuously recorded throughout the experiment using the BIOPAC AcqKnowledge software~\cite{biopac_acqknowledge}. All signals were collected at a raw sampling rate of 2000~Hz and exported in CSV format for offline analysis.

Physiological data were segmented into four conditions: \textit{control group}, \textit{stylized avatar}, \textit{realistic avatar}, and \textit{live video}. In the control condition, signals were recorded for 30 seconds after participants had completed a 2-minute resting period following sensor setup, serving as a baseline measure. The remaining conditions corresponded to conversational interactions with English native speakers represented by a stylized avatar, a realistic avatar, or a live video stream, respectively.

\subsection{Physiological Signal Processing and Anxiety-Related Feature Extraction}

\subsubsection{Signal Preprocessing}
All signals were preprocessed to reduce noise prior to feature extraction.

The EDA signal was first low-pass filtered with an upper cutoff frequency of 3~Hz to remove high-frequency noise. A notch filter at the power-line frequency (50~Hz) was then applied. The preprocessed signal was subsequently decomposed into tonic (skin conductance level, SCL) and phasic (skin conductance response, SCR) components.

The ECG signal was processed using a high-pass filter with a cutoff frequency of 0.5~Hz to remove baseline drift, followed by a 50~Hz notch filter to suppress power-line interference.

The PPG signal was filtered using a band-pass configuration, consisting of a high-pass filter at 0.5~Hz and a low-pass filter at 8~Hz, to retain the frequency components relevant to pulse detection.
\subsubsection{Feature Extraction}

\begin{table*}
  \caption{Physiological features extracted from EDA, ECG, and PPG signals for anxiety-related analysis.}
  \label{tab:biosignals}
  \begin{tabularx}{\linewidth}{cccl>{\centering\arraybackslash}Xc}
    \toprule
    Signal & Domain & Parameter & & Description & Unit \\
    \midrule
    \multirow{4}*{EDA} 
      & \multirow{2}*{Event} 
        & NR 
          & & Number of SCR responses per minute 
          & 1/min \\
    \cline{3-6}
      & 
        & maxR 
          & & Maximum SCR amplitude within the analysis window 
          & $\mu$S \\
    \cline{2-6}
      & \multirow{2}*{Time} 
        & MH 
          & & Mean SCR amplitude including tonic component 
          & $\mu$S \\
    \cline{3-6}
      & 
        & MA 
          & & Mean SCR amplitude excluding tonic component 
          & $\mu$S \\
    \hline
    \multirow{3}*{ECG} 
      & \multirow{2}*{Time} 
        & mHR 
          & & Mean heart rate 
          & 1/min \\
    \cline{3-6}
      & 
        & RMSSD 
          & & Root mean square of successive NN interval differences 
          & ms \\
    \cline{2-6}
      & Frequency 
        & LF/HF 
          & & Ratio of low- to high-frequency power 
          & -- \\
    \hline
    PPG 
      & Time 
        & MPR 
          & & Mean pulse rate derived from PPG 
          & 1/min \\
  \bottomrule
\end{tabularx}
\end{table*}

We extracted a total of eight physiological features associated with autonomic arousal and anxiety from the preprocessed EDA, ECG, and PPG signals. Table~\ref{tab:biosignals} summarizes the extracted features, their domains, and descriptions.

All signal preprocessing and feature extraction were conducted using the \textit{NeuroKit2} Python package~\cite{makowski2021neurokit2}. To ensure comparability across conversational conditions with varying durations, feature extraction was restricted to fixed-length time windows. All conversations lasted longer than two minutes; to control for variability in interaction length, analyses were restricted to features extracted from the first 120 seconds of each interaction. For the baseline control condition, only the first 30 seconds were analyzed. All selected features were defined to be independent of signal duration.

\textbf{EDA Features.}
From the EDA signal, we extracted standard SCR-based features to capture autonomic arousal. These included the number of SCR responses (NR) per minutes to reflect response frequency, the maximum SCR amplitude (maxR) to characterize peak response magnitude, and two mean amplitude measures capturing overall response intensity with (MH) and without (MA) tonic components. Together, these features represent both the frequency and magnitude of phasic electrodermal responses associated with anxiety-related arousal~\cite{Petrescu2020IntegratingBM, Makowski2021NeuroKit2AP}.

\textbf{ECG Features.}
From the ECG signal, we extracted heart rate (HR) and heart rate variability (HRV) features commonly used to reflect anxiety-related autonomic changes. Mean heart rate (mHR) was computed to capture overall cardiovascular arousal. In addition, the root mean square of successive differences (RMSSD) was extracted as an index of vagally mediated HRV. We also computed the low-frequency (0.04 - 0.15Hz) to high-frequency (0.15 - 0.4Hz) power ratio (LF/HF) to characterize the balance between sympathetic and parasympathetic activity~\cite{Petrescu2020IntegratingBM, Miranda2014AnxietyDU, Kreibig2010AutonomicNS, Shaffer2014AHH, Shaffer2017AnOO}.

\textbf{PPG Features.}
From the PPG signals, we extracted mean pulse rate (MPR) to characterize cardiovascular activity based on peripheral blood volume changes. MPR served as a complementary measure to ECG-derived heart rate, reflecting anxiety-related physiological arousal.

Research focusing on the relationship between physiological signal features and anxiety level suggests that, anxiety can result in increased mHR~\cite{Petrescu2020IntegratingBM, Miranda2014AnxietyDU, Wen2020TowardCA, diagnostics12081794}, increased LF/HF~\cite{Pham2021HeartRV,diagnostics12081794,Perpetuini2021PredictionOS}, increased NR~\cite{Petrescu2020IntegratingBM,diagnostics12081794}, increased maxR~\cite{Petrescu2020IntegratingBM, Lee2020DetectionOD}, increased MH~\cite{Petrescu2020IntegratingBM,Lee2020DetectionOD}, increased MA~\cite{Petrescu2020IntegratingBM,Lee2020DetectionOD,diagnostics12081794}, increased MPR~\cite{Lee2020DetectionOD,diagnostics12081794} and also decreased RMSSD~\cite{Malik1996HeartRV,Stein1994HeartRV,diagnostics12081794,Perpetuini2021PredictionOS}.

\section{Results}
We separately analyzed anxiety as reflected in physiological measures and self-report survey responses, and further examined avatar visual characteristics associated with anxiety based on qualitative feedback from participants.

\subsection{Physiological Indicators of Anxiety Across Representation Types}
\begin{figure}[ht]
\centering
\includegraphics[width=\columnwidth]{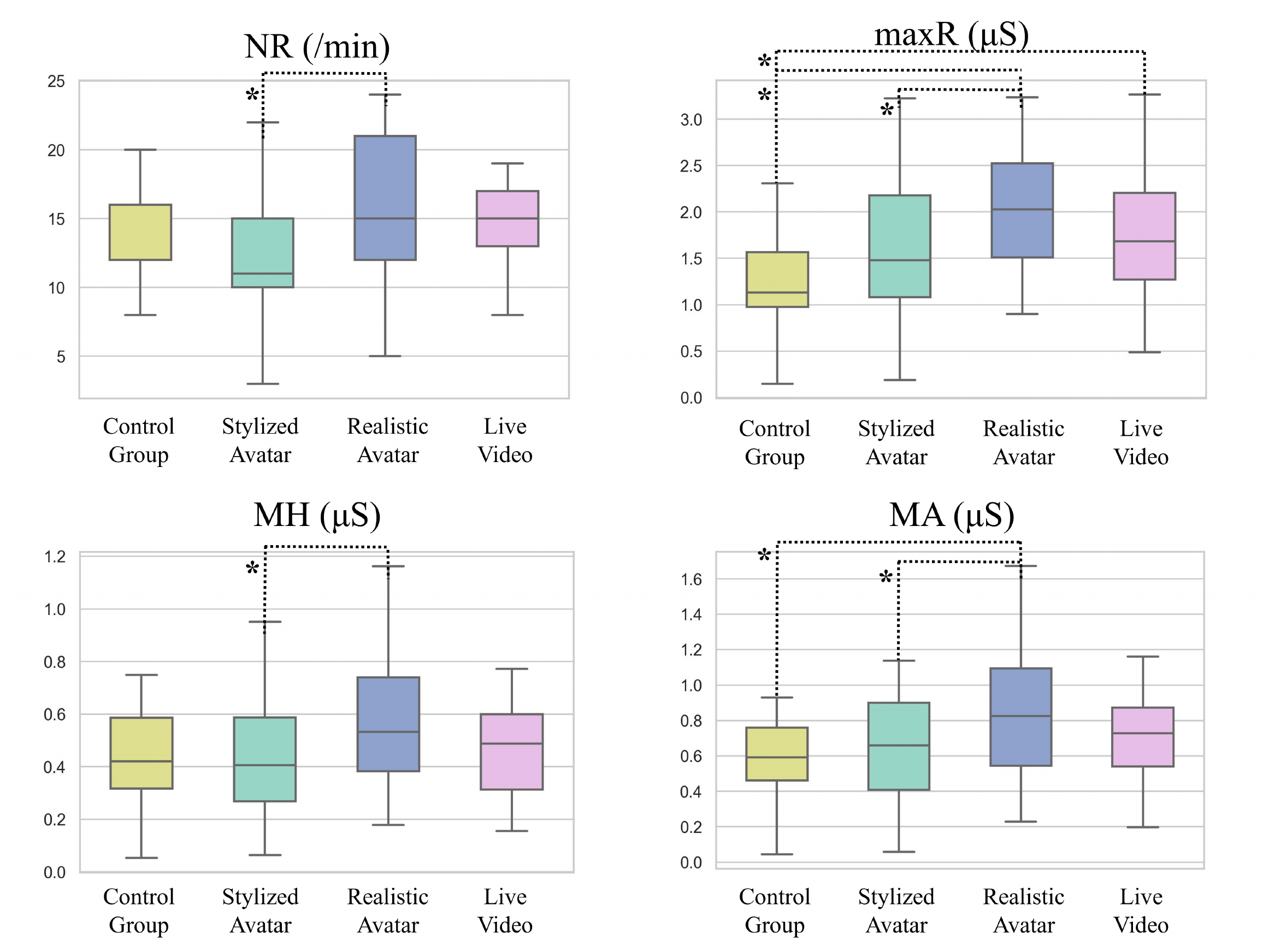}
\caption{Distributions of NR, maxR, MH, and MA features when participants were in the control group or interacting with stylized avatars, realistic avatars, or live videos. Dotted lines marked with an asterisk (*) highlight significant differences between paired groups (p < 0.05). The central line in each box represents the median.}
\label{fig_feature_distribution}
\end{figure}

We analyzed physiological features associated with user anxiety, including NR, maxR, MH, MA, mHR, RMSSD, LF/HF, and MPR. Significant effects were observed in NR, maxR, MH, MA features. 

\textbf{NR Analysis.} Shapiro--Wilk tests indicated that NR did not meet the normality assumption across conditions. A Friedman test revealed a significant effect of condition on NR, $\chi^2(3)=9.452$, $p=.024$. Bonferroni-adjusted post hoc comparisons identified a significant difference between the stylized avatar and realistic avatar conditions ($p=.037$).

\textbf{maxR Analysis.} Shapiro--Wilk tests indicated that the normality assumption for maxR was not violated. The sphericity assumption was met. A repeated-measures ANOVA revealed a significant effect of condition on maxR, $F(3,72)=10.788$, $p<.001$. Bonferroni-adjusted post hoc comparisons showed significant differences between the Control and realistic avatar conditions ($p<.001$), the Control and live video conditions ($p=.036$), and the stylized avatar and realistic avatar conditions ($p=.018$).

\textbf{MH Analysis.} Shapiro--Wilk tests indicated that MH did not meet the normality assumption. A Friedman test revealed significant differences in MH across conditions, $\chi^2(3)=8.856$, $p=.031$. Bonferroni-adjusted pairwise comparisons showed a significant difference between the stylized avatar and realistic avatar conditions ($p=.019$).

\textbf{MA Analysis.} Shapiro--Wilk tests indicated that MA did not meet the normality assumption. A Friedman test revealed a significant effect of condition on MA, $\chi^2(3)=14.472$, $p=.002$. Bonferroni-adjusted post hoc comparisons indicated significant differences between the realistic avatar and Control conditions ($p=.006$) and between the realistic avatar and stylized avatar conditions ($p=.006$).

No significant differences were found for mHR, RMSSD, LF/HF, MPR features ($p > 0.05$). The distributions of features showing significant effects are illustrated in Figure~\ref{fig_feature_distribution}. 

In general, interactions with the realistic avatar were associated with higher levels of physiological anxiety-related arousal across multiple features compared to the stylized avatar. In contrast, the live video condition showed relatively lower levels of arousal. In particular, the stylized avatar condition exhibited similarly low physiological responses, with no significant differences observed between the stylized avatar and live video conditions. In fact, the median values of NR, maxR, MH, and MA were slightly lower in the stylized avatar condition than in live video.

These findings suggest that, in contrast to realistic avatars, which may elevate anxiety-related physiological responses, stylized avatars provide a communication experience comparable to live video in terms of physiological anxiety and may even offer slight advantages in reducing anxiety-related physiological arousal.

\subsection{Subjective Anxiety Across Representation Types}

To measure ESL speakers' anxiety during the three communication representations, we administered a within-study survey after each exposure: stylized avatar (C1), realistic avatar (C2), and live video (C3). Table~\ref{tab:avatar-video-items} lists the survey items (E-S1--E-S6). Responses were aggregated into a composite score by averaging E-S1--E-S6, where higher scores indicate lower anxiety.

\begin{table}[t]
  \caption{Survey items measuring participants' anxiety during avatar- or video-mediated conversations.}
  \label{tab:avatar-video-items}
  \centering
  \begin{tabularx}{\columnwidth}{c X}
    \toprule
    Code & Statement \\
    \midrule
    E-S1 & I found it difficult to look at his/her face while speaking. \\
    E-S2 & Most of the time, I felt uncomfortable when looking at him/her. \\
    E-S3 & My eyes tended to wander while speaking to him/her. \\
    E-S4 & I had trouble speaking clearly while looking at him/her. \\
    E-S5 & His/her appearance distracted me during the conversation. \\
    E-S6 & I felt anxious when talking to him/her. \\
    \bottomrule
  \end{tabularx}
\end{table}

\begin{itemize}
\item C1: Anxiety when speaking to stylized avatar
\item C2: Anxiety when speaking to realistic avatar
\item C3: Anxiety when speaking to live video of real person
\end{itemize}


The anxiety scale demonstrated good internal consistency across all three representation conditions. Cronbach's alpha values were .79 for the stylized avatar condition, .80 for the realistic avatar condition, and .82 for the live video condition, supporting the aggregation of items E-S1 through E-S6 into a composite anxiety score.

Using anxiety scores, we tested whether the representation type affected the' communication anxiety of ESL speakers. A Friedman test revealed a significant effect of representation type ($\chi^2(2)=24.66$, $p<0.001$), with a large effect size (Kendall's $W=0.50$). Descriptively
, participants reported the lowest anxiety in the live video condition ($M=2.88$, $SD=1.00$), followed by the stylized avatar ($M=3.19$, $SD=0.99$), while the realistic avatar elicited the highest anxiety ($M=3.72$, $SD=1.14$).

Post-hoc Wilcoxon signed-rank tests with Benjamini–Hochberg correction indicated significant differences across all pairwise comparisons. Anxiety in the stylized avatar condition was significantly lower than in the realistic avatar condition ($q=0.009$). The live video condition was associated with significantly lower anxiety than both avatar conditions, with a smaller difference relative to the stylized avatar ($q=0.038$) and a larger difference compared to the realistic avatar ($q<0.001$).

The item-level analysis and corresponding box plots for statements E1–E6 are provided in the supplementary material, as the six items represent indicators of the same ESL communication anxiety construct.


\subsection{Characteristics in Avatar Visual Style Associated with ESL Speakers' Anxiety}


To further understand why avatar style can affect communication anxiety, we examined how specific visual characteristics of avatar representations were associated with ESL speakers' anxiety. In the post-study survey, participants were asked to compare the three visual representations from the aspect of communication anxiety and provide open-ended explanations for their communication anxiety when speaking to the avatars or live video. Participants are referred to by anonymized identifiers (P1–P25).
We then conducted a thematic analysis of participants' open-ended explanations, following the six-phase approach proposed by Braun and Clarke~\cite{braun2006using}. Two researchers independently coded the responses, iteratively discussed discrepancies, and refined themes through consensus. The analysis focused on identifying recurring patterns related to facial expressiveness and social cues, perceived feedback and fear of negative evaluation, and contextual appropriateness.

\subsubsection{Facial Expressiveness and Social Cues}
Across conditions, participants consistently emphasized the importance of facial expressiveness and coherent social cues to take control of anxiety. Live video was widely valued for its rich and natural facial dynamics, including eye contact, emotional expressions, and fine-grained mouth movements. These cues were reported as critical for interpreting the interlocutor’s reactions and sustaining conversational flow. 

For ESL speakers, anxiety can be caused by linguistic and social uncertainty. Participants described relying facial emotional expressions to infer feedback and evaluate how well they were being understood, \textit{"I want to see the other person's expressions and get more information... Otherwise, I feel higher anxiety because I don't know if I expressed myself correctly"} (P5). At the same time, mouth shape was highlighted as especially important for language comprehension, as it provides additional phonetic information that supports understanding, \textit{"Facing a real person, mouth shape helps me follow her pace"} (P8). These findings suggest that different aspects of facial information serve distinct functions for ESL communication: emotional expressions help assess social feedback, while mouth movements support linguistic processing.

In contrast, both stylized and realistic avatars were frequently described as providing reduced and less reliable facial information compared to live video. Participants noted limited mouth movement, less dynamic expressions, and less natural eye behavior, which constrained their ability to interpret both emotional feedback and speech-related cues. This reduction in non-verbal cues was associated with increased uncertainty and, in some cases, elevated anxiety during interaction.

\subsubsection{Perceived Feedback and Fear of Negative Evaluation}

Fear of negative evaluation is a key factor in anxiety of ESL speakers~\cite{horwitz1986foreign}. Stylized avatars were consistently associated with lower perceived negative feedback and reduced fear of negative evaluation. As one participant noted, \textit{"This avatar is my favorite because it looks so nice and friendly"} (P22), the perceived negative effect was lower when speaking to stylized avatar. Some participants attributed this effect to lower fear of negative evaluation by the stylized interlocutor, and they experienced a reduced tendency to over-monitor the interlocutor's reactions, as reflected in the comment: \textit{"I don't worry much about my grammar being judged (by the stylized avatar)."} (P17). As a result, stylized avatars were perceived as creating a lower-pressure interaction environment for ESL participants.

Responses to live video were mixed. Many participants found interaction with a real person reassuring, as it allowed access to clear positive feedback. For example, \textit{"Talking to a real person feels less stressful because I can get positive feedback from them"} (P11). However, some participants reported higher anxiety in live video compared to stylized avatars, describing increased self-awareness and "stage fright" (P17).

Realistic avatars were most frequently associated with higher levels of anxiety. One possible explanation is the \textit{uncanny valley}, often described as a level of realism at which avatars evoke feelings of eeriness due to subtle visual abnormalities~\cite{nowak2018avatars}. However, this explanation alone does not fully account for our findings. Only a small number of participants (2/25) explicitly described the realistic avatars as “creepy” or “frightening,” suggesting that visual eeriness was not the primary driver of increased anxiety.

Instead, our findings are more consistent with an expectation-based interpretation of the uncanny valley, in which visually realistic representations are associated with heightened expectations for social feedback~\cite{hamilton2010advancing}. Prior work suggests that users anticipate richer and more natural responses when interacting with realistic avatars~\cite{nowak2018avatars}. 

In the ESL context, where users are often uncertain about their own performance, timely and socially responsive feedback from an interlocutor can help reduce this uncertainty. Like P5 said, \textit{"Seeing their reaction helps"}. When visual feedback is limited, users may experience increased uncertainty about how they are being perceived, which amplifies perceived negative evaluation and fear of negative evaluation, contributing to higher communication anxiety. In line with this observation, P8 mentioned that, \textit{"Compared to the cute one (the stylized avatar), this one (the realistic) looks quite realistic, but it does not provide lively facial feedback... makes the person feel hard to engage with"}. 

\subsubsection{Contextual Appropriateness}

Perceptions and expectations of contextual appropriateness were also related to anxiety. Live video was generally perceived as the most appropriate for formal settings, such as lectures and conferences, or interaction between acquaintances. 
In contrast, stylized avatars were commonly described as better suited for casual contexts, such as language practice with a study partner, cross-cultural social discovery, or entertainment-oriented settings (e.g., games), particularly when communicating with strangers. In these contexts, stylized avatars were perceived as reducing anxiety (e.g., \textit{"If this were a game setting, I would like to talk to this (stylized avatar) instead of the camera"} (P22)). Realistic avatars, positioned between stylized avatars (informal) and live video (formal and familiar), were often associated with ambiguity in social norms. This lack of clear contextual fit introduced uncertainty about appropriate norms, which participants reported as contributing to increased anxiety.

These observations point to the possible role of contextual appropriateness in shaping user perceptions. However, the current context was designed as a semi-formal conversation based on prior work~\cite{yoneyama2023augmented, mcdonnell2012render}. As context was not systematically manipulated in this study, these insights should be interpreted as exploratory and warrant further investigation.

\section{Discussion}

Our results show that avatar style plays an important role in shaping communication anxiety for ESL speakers. Live video remained a strong baseline, consistently associated with low anxiety in both physiological and self-reported measures. At the same time, stylized avatars were associated with significantly lower physiological anxiety than realistic avatars and exhibited anxiety levels comparable to—and even slightly lower than—live video. From subjective feedback, some participants showed a modest preference for live video, likely due to its richer facial cues and contextual familiarity, while others noted that stylized avatars were associated with reduced perceived negative feedback and fear of evaluation and were considered more suitable for less formal or entertainment-oriented contexts. In contrast, realistic avatars were associated with elevated anxiety. 

The observed differences can be understood through three interrelated factors: expressive richness, perceived negative feedback, and contextual appropriateness.
Live video provides rich and reliable social signals, including subtle facial expressions and mouth movements, which can help reduce uncertainty and anxiety during communication. The animation tools used in our study are representative of those commonly used for 2D avatar control in recent years ~\cite{guajardo2024generative, willett2020pose2pose}. Although avatars attempt to replicate these cues, current avatar control techniques are limited in their ability to convey nuanced nonverbal signals.

At the same time, stylized avatars appear to offer a complementary advantage. Their lower-fidelity, more abstract appearance may reduce users’ expectations of precise, immediate social feedback. As a result, when the feedback is ambiguous or limited, users may be more tolerant of such imperfections and less likely to interpret them as negative evaluations. This can help reduce the perceived pressure to be judged and, in turn, reduce communication anxiety. Participant feedback also suggests that stylized avatars may be particularly suitable for more casual or entertainment-oriented contexts, where strict social expectations are less prominent.

Together, these findings suggest several implications for avatar design in ESL communication.

First, adopting stylized representations with lower visual realism—such as exaggerated facial features and familiar cartoon-like designs—may help reduce communication anxiety through various psychological mechanisms.

Second, individual differences highlight the value of personalization. Although general trends were observed, participants varied in their preferences: Some preferred stylized avatars to reduce anxiety, whereas others preferred live video for its familiarity and clarity. Providing users with options to adjust or switch representations may better accommodate diverse needs and preferences.

In addition, context may influence how different representations are perceived. Our subjective findings suggest that stylized avatars may be more suitable for informal or entertainment-oriented interactions, whereas live video remains advantageous in contexts that require higher levels of social presence and familiarity. As the context was not manipulated as an independent variable in this study, these observations should be interpreted with caution. However, they point to promising directions for future work on expressive avatar design.

\subsection{Limitations and Future Work}
This study has several limitations that point to important directions for future work.

First, our participant sample was linguistically homogeneous, consisting exclusively of Mandarin-speaking ESL users. Given the diversity of ESL populations worldwide, cultural and linguistic backgrounds can shape how social cues, evaluation, and visual representations are perceived. As a result, our findings should be interpreted as most directly applicable to Mandarin ESL speakers, and caution is warranted when generalizing to ESL users from other languages or cultural backgrounds. Future studies should examine whether the observed patterns hold across diverse ESL populations.


Second, ESL communication occurs across a wide range of contexts, yet our study focused on an interview-based conversation setting; following the previous work~\cite{yoneyama2023augmented, mcdonnell2012render}. While this context provides a meaningful and moderately high-stakes interaction, it does not represent the full spectrum of ESL communication scenarios. Future research should investigate how avatar styles influence anxiety across different contexts, such as casual conversations, collaborative tasks, and public speaking.

Third, the distinction between stylized and realistic avatars represents a broad conceptual categorization rather than a precise boundary. Perceptions of realism likely exist along a continuum, and what constitutes an appropriate or comfortable level of realism may vary across individuals and contexts. Further research could systematically parameterize avatar realism to investigate at what point increases in realism begin to introduce discomfort for ESL speakers.


\section{Conclusion}

This study explored how avatar style influences communication anxiety among ESL speakers, using both physiological signals and subjective feedback, and compared these effects with live video. ESL speakers are a key focus, as their communication experiences are often strongly shaped by anxiety.

Overall, our findings suggest that, from a physiological perspective, stylized avatars can match—and in some cases slightly outperform—live video in terms of communication anxiety, offering a promising alternative representation for ESL communication. Meanwhile, subjective feedback shows that many participants still preferred live video, primarily due to its richer facial cues. At the same time, participants reported perceiving a lower negative social evaluation when using stylized avatars and highlighted their potential in more casual or entertainment-oriented contexts. 

This work provides a step forward towards understanding how avatar representations can support emotionally sustainable cross-linguistic communication. Although we focus on ESL speakers as a representative population experiencing communication anxiety, these findings may offer broader insights for avatar-mediated interaction, where reducing social anxiety while maintaining effective communication remains a key challenge.

\bibliographystyle{ACM-Reference-Format}
\bibliography{sample-base}





\end{document}